\documentclass[%
reprint,
superscriptaddress,
aps
amsmath,amssymb,
reprint,%
floatfix,
showkeys,
]{revtex4-2}

\usepackage{graphicx}
\usepackage{dcolumn}
\usepackage{bm}
\usepackage[normalem]{ulem}
\usepackage{mathptmx}
\usepackage{color}
\usepackage{hyperref}
\usepackage{orcidlink}
\usepackage[mathlines,pagewise]{lineno}
\relax 

\begin{document}


\title[]{Hall mobilities and sheet carrier densities in a single LiNbO$_3$ conductive ferroelectric domain wall}

\author{Henrik Beccard\, \orcidlink{0000-0001-9331-1294}}%
\affiliation{Institute of Applied Physics, Technische Universit\"at Dresden, N\"othnitzer Strasse 61, 01187 Dresden, Germany}%
\author{Elke Beyreuther\, \orcidlink{0000-0003-1899-603X}} 
\email{elke.beyreuther@tu-dresden.de}%
\affiliation{Institute of Applied Physics, Technische Universit\"at Dresden, N\"othnitzer Strasse 61, 01187 Dresden, Germany}%
\author{Benjamin Kirbus\, \orcidlink{0000-0002-8824-2244}}%
\affiliation{Institute of Applied Physics, Technische Universit\"at Dresden, N\"othnitzer Strasse 61, 01187 Dresden, Germany}%
\author{Samuel D. Seddon\, \orcidlink{0000-0001-8900-9308}}
\affiliation{Institute of Applied Physics, Technische Universit\"at Dresden, N\"othnitzer Strasse 61, 01187 Dresden, Germany}%
\author{Michael R\"using\,\orcidlink{0000-0003-4682-4577}}%
\affiliation{Institute of Applied Physics, Technische Universit\"at Dresden, N\"othnitzer Strasse 61, 01187 Dresden, Germany}%
\affiliation{Integrated Quantum Optics, Institute for Photonic Quantum Systems (PhoQS), Paderborn University, 33098 Paderborn, Germany}
\author{Lukas~M.~Eng\,\orcidlink{0000-0002-2484-4158}}
\affiliation{Institute of Applied Physics, Technische Universit\"at Dresden, N\"othnitzer Strasse 61, 01187 Dresden, Germany}%
\affiliation{ct.qmat: Dresden-W\"urzburg Cluster of Excellence--EXC 2147, Technische Universit\"at Dresden, 01062 Dresden, Germany}%

\date{\today}

\begin{abstract}
For the last decade, conductive domain walls (CDWs) in single crystals of the uniaxial model ferroelectric lithium niobate (LiNbO$_3$, LNO) have shown to reach resistances more than 10 orders of magnitude lower as compared to the surrounding bulk, with charge carriers being firmly confined to sheets of a few nanometers in width. LNO thus currently witnesses an increased attention since bearing the potential for variably designing room-temperature nanoelectronic circuits and devices based on such CDWs. In this context, the reliable determination of the fundamental transport parameters of LNO CDWs, in particular the 2D charge carrier density $n_{2D}$ and the Hall mobility $\mu_{H}$ of the majority carriers, are of highest interest. In this contribution, we present and apply a robust and easy-to-prepare Hall-effect measurement setup by adapting the standard 4-probe van-der-Pauw method to contact a single, hexagonally-shaped domain wall that fully penetrates the 200-$\mu$m-thick LNO bulk single crystal. We then determine $n_{2D}$ and $\mu_{H}$ for a set of external magnetic fields $B$ and prove the expected cosine-like angular dependence of the Hall voltage. Lastly, we present photo-Hall measurements of one and the same DW, by determining the impact of super-bandgap illumination on the 2D charge carrier density $n_{2D}$. 
\end{abstract}


\maketitle

\section{\label{sec:Intro}Introduction}
 
Continuous progress in solid-state nanotechnology relies on answering a number of unsolved scientific questions with respect to both material systems and device operation principles. One such burning issue is electrical transport under well-defined and controlled conditions in reduced dimensions, as are the 2-dimensional (2D) material systems. While, for example, the 2D van-der-Waals materials are thoroughly analyzed~\cite{nov16}, our focus here lies on 2D sheets built up from conductive ferroelectric domain walls (CDWs), i.e., the transition regions between ferroelectric domains of opposite dielectric polarization, which can be tuned to exhibit a strongly enhanced conductivity as compared to the surrounding bulk \cite{god17,kir19}. In fact, DWs in ferroelectrics have been reported to form effective 2D electron gases (2DEGs) after applying specialized preparation routines to these wide-bandgap bulk materials~\cite{slu13,bec22}. LiNbO$_3$ (LNO) turned out to be \textit{the} "drosophila" ferroelectric for CDW engineering, since it is robust, semiconductor compatible, and easy-to-reconfigure at room temperature, while being commercially available both as a bulk material and in crystalline thin-film-on-insulator form \cite{Rusing2019b,Zhu2021,Sun2020,Kampfe2020}. Notably, this ever rising interest in CDWs has been reviewed with respect to both theoretical and device oriented aspects in a number of excellent works \cite{cat12,mei15,slu16,bed18,sha19,nat20,mei21,sha22}. 

At the fundamental level, primarily, the divergence of the ferroelectric polarization (i.e., the local vector field that describes the volume density of unit cell dipoles) at the DW has been identified as one of the main driving forces behind the localized DW conductivity, that, not to forget, has been predicted already in the 1970s~\cite{vul73}. This divergence is understood as an intrinsic charge density acting as the source of the so-called depolarization field. In turn, the emergent electrostatic field leads to the attraction of free charge carriers to the DWs, as well as to the population of otherwise free electron and/or hole states due to local band bending at the DW position. For the simplest case of a uniaxial ferroelectric (as is LNO) that shows purely Ising-type DWs, this divergence is directly related to the geometrical inclination of the DW with respect to the polar axis \cite{eli11,god17}.

\begin{figure*}[ht]
\includegraphics[width=\textwidth]{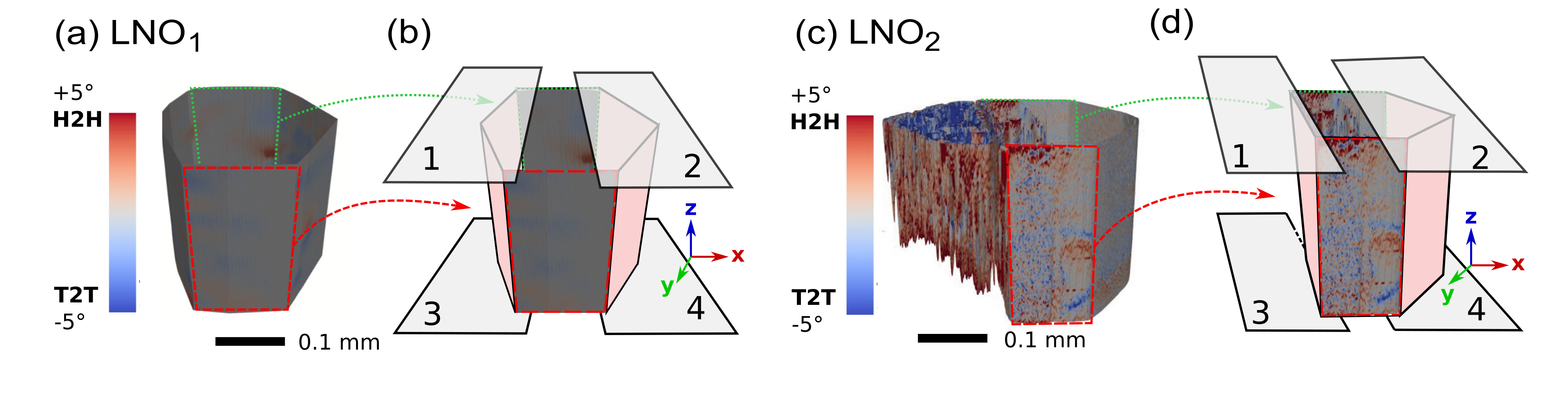}
\caption{\label{fig:1} (a,c) 3D Cherenkov second-harmonic generation microscopy (CSHG) data, and (b,d) chromium electrode arrangement 1--4 for the two samples "LNO$_1$" and "LNO$_2$", as prepared from z-cut LiNbO$_3$ single crystals for Hall-transport measurements using the van-der-Pauw method. Note that contacts 1--4 inclose a parallel junction of 2 conductive domain walls (CDWs), one in the front (red, dashed) and one in the back (green, dotted). The different protocols applied for domain-wall conductivity (DWC) enhancement in "LNO$_1$" (a,b) and "LNO$_2$" (c,d) resulted in the different shape and appearance, as seen in (a,c). The enhanced and desired head-to-head ("H2H") DW inclination is color coded in red in the CSHG scale bar, while  tail-to-tail type DWs ("T2T") appear in blue. As seen, sample "LNO$_2$" shows a larger DW inclination enhancement, justifying the larger DW current as compared to sample LNO$_1$.}
\end{figure*}

Nevertheless, on a practical level, the determination of DW-related quantitative transport data such as the charge carrier type, density, and mobility, was, to date, restricted to a few exemplary cases, only. In this context, especially the analysis of the Hall effect in DWs in the improper ferroelectrics YbMnO$_3$ and ErMnO$_3$ has shown to be a valuable tool, as reported in the groundbreaking papers of Campbell \textit{et al.}~\cite{cam16} and Turner \textit{et al.}~\cite{tur18}, respectively. Those authors chose scanning-probe-based approaches for their evaluation, needing a cumbersome and sophisticated procedure to disentangle the Hall potential from the cantilever-based three-terminal reading via calibration routines and accompanying simulations. Moreover, as these authors state themselves, their approach is mainly valid to extract near-surface charge carrier densities and mobilities, only. With respect to proper ferroelectrics, the Hall-effect investigation by Qian \textit{et al.}~\cite{qia22} of DW pn-junctions engineered into x-cut thin-film lithium niobate (TFLN), was equally limited to surface-near carrier densities, while McCluskey \textit{et al.}~\cite{mccluskey_ultrahigh_2022} made use of the fact that a DW in z-cut TFLN mirrors the Corbino cone geometry and in turn found promisingly high carrier mobilities, which were extracted from a magnetoresistance analysis. Notably, the recent study by Beccard \textit{et al.}~\cite{bec22} proposes a completely different approach, quantifying both the 2D charge carrier densities $n_{2D}$ and Hall mobilities $\mu_{H}$ through "macroscopic" Hall-effect measurements, by adapting the classical van-der-Pauw (vdP)~\cite{vdP58} four-point electrode configuration to measurements from a single CDW in bulk BaTiO$_3$. 

The work presented in this paper sets in exactly at this point, by adopting the vdP scenario to the particular case of a single CDW in z-cut bulk LiNbO$_3$, the uniaxial model ferroelectric of uttermost importance for prospective nanoscale applications. While significantly high DW conductivity (DWC) in LNO has been proven for the last decade in a number of consecutive works \cite{sch12,wer17,god17,kir19}, Hall-effect measurements in LNO CDWs are still lacking. In fact, the challenge consists in adapting the vdP method to the hexagonally-shaped DW in LNO as depicted in Fig.~\ref{fig:1}; as seen, the four vdP contacts then likely measure the parallel junction of two such conductive DWs in LNO, rather than connecting to one single planar 2D sheet as was the case for the CDW in BaTiO$_3$~\cite{bec22}. Nonetheless, in the following we will show that evaluating the DW sheet resistance and the magnetic-field dependent Hall voltages still allows applying the vdP method, hence revealing quantitative data for both $n_{2D}$ and $\mu_{H}$ on a so far unprecedented level and precision. In fact, the error stemming from the parallel DW junction is a factor of two in maximum, as easily figured when calculating the total resistance of a parallel junction of two identical / different resistors, i.e., two CDWs. Notably, this factor of two does not change the order of magnitude of both $n_{2D}$ and $\mu_{H}$. In addition, the same error factor of two is obtained when analyzing the Hall-data in a more rigorous way by applying the concept of the resistor network (RN)~\cite{wol18}, as is discussed in the SI-section~D. 

Moreover, the integrity of the results obtained by the vdP method is corroborated here by two further investigations, (i) by quantifying the angular Hall-voltage dependence, and (ii) when inspecting the Hall-voltage response under super-bandgap illumination for purposely generating additional electron-hole pairs within the CDW.

\section{Materials and Methods}

\subsection{Samples -- Fabrication of Domain Walls}

For our study here, we employed two 5-mol\% Mg-doped congruent z-cut LiNbO$_3$ crystal plates, with sizes of approx.~1$\times$0.5~mm$^2$ in the xy-plane and a thickness of 200~\textmu m in the polar z-direction, cut from a commercial wafer by Yamaju Ceramics Co., Ltd.. In the following, we label these samples as "LNO$_1$" and "LNO$_2$". A single, fully penetrating and hexagonally-shaped ferroelectric domain of a 257~\textmu m~(LNO$_1$) and 356~\textmu m~(LNO$_2$) diameter [SI-Figs.~S7(a) and (b)] was then grown into these samples by applying the well-established method of UV-assisted poling~\cite{wen05,sch12} using liquid electrodes and a He-Cd laser (Kimmon Koha IK3301R-G) operated at a 325~nm wavelength, for more details refer, e.g., to~\cite{sch12,kis23}. Then, four 8-nm-thick chromium electrodes were vapor-deposited under high-vacuum conditions onto every DW structure, using a shadow mask. The exact electrode geometry and arrangement with respect to the two crystals and DW orientations are sketched in Fig.~\ref{fig:1}(b,d), while a polarization-sensitive microscopy top-view image is found in the SI-Fig.~S7(c). In the following, the electrodes are consecutively labelled by indices 1--4, as standard to 4-point van-der-Pauw experiments. As a result, the 4 vdP electrodes 1--4 directly contact to the two DWs that lie in the xz-plane, one in the front (red, dashed), one in the back (green, dotted), as seen in Fig.~\ref{fig:1} for samples "LNO$_1$" and "LNO$_2$", respectively. Note that in parallel, two monodomain reference samples "LNO$_3$" and "LNO$_4$" were prepared as well, having identical electrodes 1 -- 4, but containing no DWs.

\begin{figure}[h]
\includegraphics[width=0.48\textwidth]{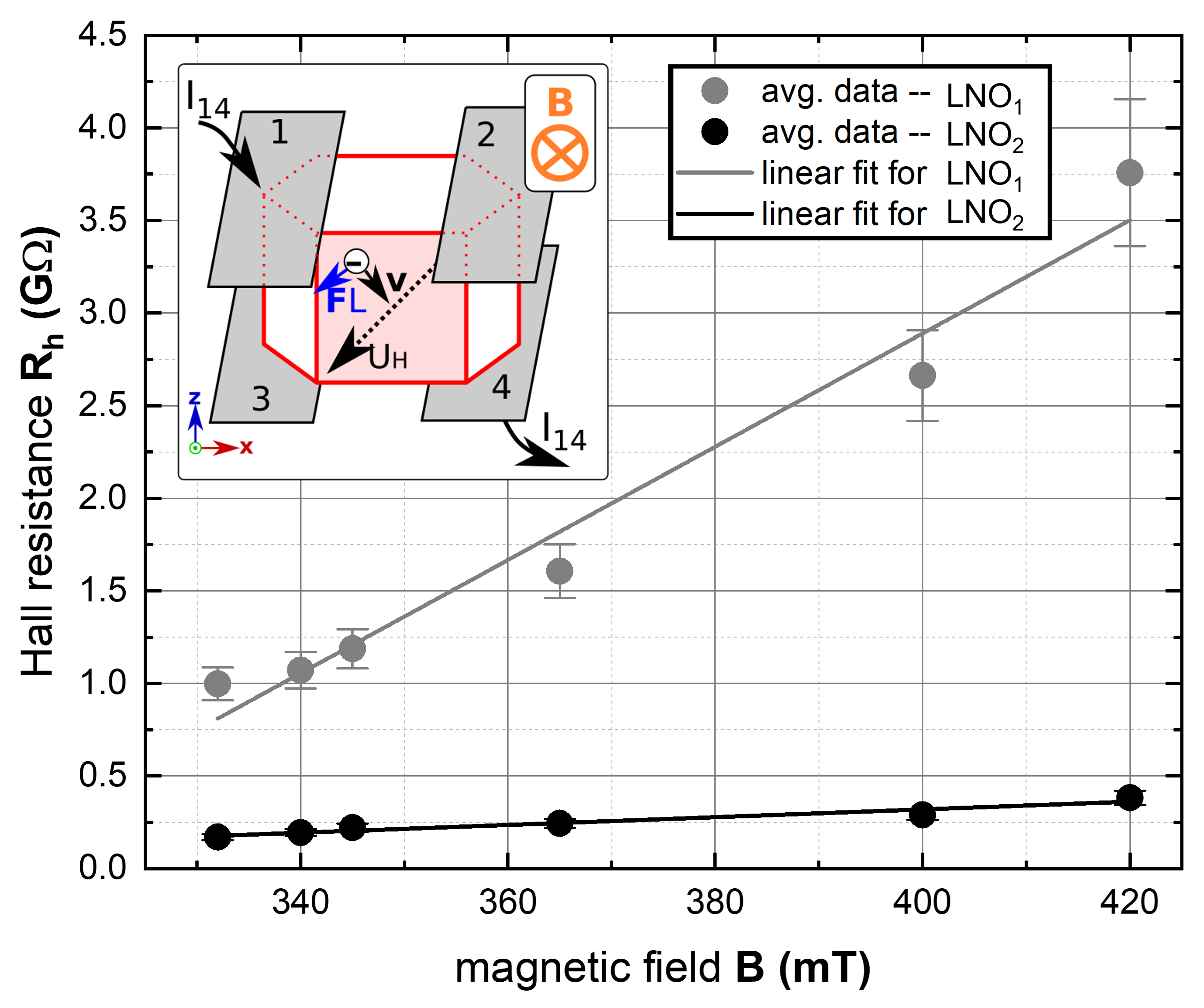}
\caption{\label{fig:2} Results of the macroscopic Hall-effect measurements for samples LNO$_1$ and LNO$_2$, carried out as sketched by the inset. The $B$-field was applied perpendicular to the CDWs and then swept both positively and negatively. The charge carriers, i.e., mainly electrons -- as discussed in the text -- that flow between the current contacts 1 and 4 are deflected by the Lorentz force $F_L$ resulting in a measurable Hall voltage $U_H = U_{23}$ read between contacts 2 and 3. Plotted in the main diagram is the Hall resistance $R_h = U_{23}/I_{14}$ as a function of $B$-field for the two samples, with sample LNO$_2$ showing a nearly 10-times larger response. Note that the plotted $R_h$ values are the averaged values of $R_h(+B)$ and $R_h(-B)$ in order to compensate for sample misalignment effects [see raw data in SI-Fig.~S5, and the reference $U_H$ measurement recorded from a monodomain bulk LNO crystal (containing no CDW at all) in SI-Fig.~S6(a)]. The corresponding 2D charge carrier densities $n_{2D}$ were then extracted from the slope of these linear relationships according to eq.~(\ref{eq:2D_carrier_density}) and are discussed further down as well as summarized in Tab.~\ref{tab:summary_calculations}.}
\end{figure}

\subsection{Enhancement of the Domain Wall Conductivity}

Subsequently, the as-grown hexagonal DWs in samples "LNO$_1$" and "LNO$_2$" underwent the DWC "enhancement" procedure by applying high voltages between the z+ and z- sides supplied by the voltage source of a Keithley 6517B electrometer, as described previously in refs.~\cite{god17,kir19}. The corresponding current-voltage curves recorded during these voltage ramps are depicted in SI-Fig.~S1, and the exact parameters of the post-growth treatment are stated in the SI-tab.~S1. Our enhancement procedure leads to higher average DW inclination angles relative to the polar z-axis and, in turn, to stronger DW confined charge accumulation and thus to a larger DW conductivity. However, in order to possibly break up the initial parallel DW junction arrangement [see again Fig.~\ref{fig:1}] and to apply our Hall-effect measurements to a single CDW, the enhancement procedure here was realized differently and deliberately asymmetrical for the two LNO samples. In particular, we treated sample LNO$_1$ with only one high-voltage ramp between electrodes 1 and 3, while sample LNO$_2$ experienced consecutive voltage ramps between both top-bottom electrode pairs 1/3 and 2/4, respectively, cf.~SI-tab.~S1 and SI-Fig.~S1. Accompanying images by 3D Cherenkov second-harmonic generation microscopy (CSHG), our standard non-destructive and real-space method of choice for both visualizing ferroelectric DWs~\cite{kaem14,kaem15,kir19} and especially for correlating their local inclination to the DW conductivity, indeed elucidates a very different DW appearance for samples LNO$_1$ and LNO$_2$ [see Fig.~\ref{fig:1}(a) and (c)] with average inclination angles between  0$\symbol{23}$ and 0.5$\symbol{23}$ and broad inclination distributions. The two different enhancement procedures for samples LNO$_1$ and LNO$_2$ are clearly reflected in the two CSHG pictures in Fig.~\ref{fig:1}(a),(c), where the walls in LNO$_2$ show on average a much larger distribution of angles suggesting a larger local screening charge, which is later reflected in the carrier densities; nonetheless, the overall DW conductivities could be readily enhanced for both cases, by three (LNO$_1$) and six (LNO$_2$) orders of magnitude in maximum (cf. SI-Fig.~S2(a) and (b) and SI-tab.~S1). The detailed current-voltage characteristics recorded between the different electrode pairs can be found in the Supplemental Information, part A.

\subsection{Realization of Hall Voltage Measurements}

For quantifying the LNO DW Hall voltage, the adapted vdP configuration~\cite{vdP58} was employed as illustrated in the inset of Fig.~\ref{fig:2} and, within a previous study, successfully tested for 2DEGs in BaTiO$_3$ CDWs~\cite{bec22}. The sample therefore was mounted into an electromagnet at room temperature that delivers magnetic fields $B$ of up to $\pm$420~mT. Contacts 1 and 4 were connected to the Keithley 6517B electrometer to apply a bias voltage of 6~V, which resulted in a domain wall current $I=I_{14}$ on the order of 0.1~nA. The corresponding carriers hence experience the Lorentz force $F_L$ as sketched in the inset of Fig.~\ref{fig:2}, resulting in the Hall voltage $U_H:=U_{23}$ that is detected between contacts 2 and 3 using a Keithley 2700 multimeter. The ratio $R_h=U_H/I$, subsequently denoted as the Hall resistance, was determined for six different $B$-field values set between 330~mT and 420~mT. In order to account for any sample misalignment within the electromagnet (i.e. non-parallel alignment of $B$-field and DW normal vector), the $B$-field direction was switched by changing the sign of the electromagnet's voltage, and the measurement series was repeated -- a common practice for Hall voltage measurements\cite{wer17b}. The corresponding data set was acquired for both DW samples LNO$_1$ and LNO$_2$ (see Fig.~\ref{fig:2} for the averaged data and SI-Fig.~S5 for the raw data, i.e., the $B$-field-direction dependent data).

\subsection{Angular Dependence of the Hall Voltage}

\begin{figure*}[ht]
\includegraphics[width=\textwidth]{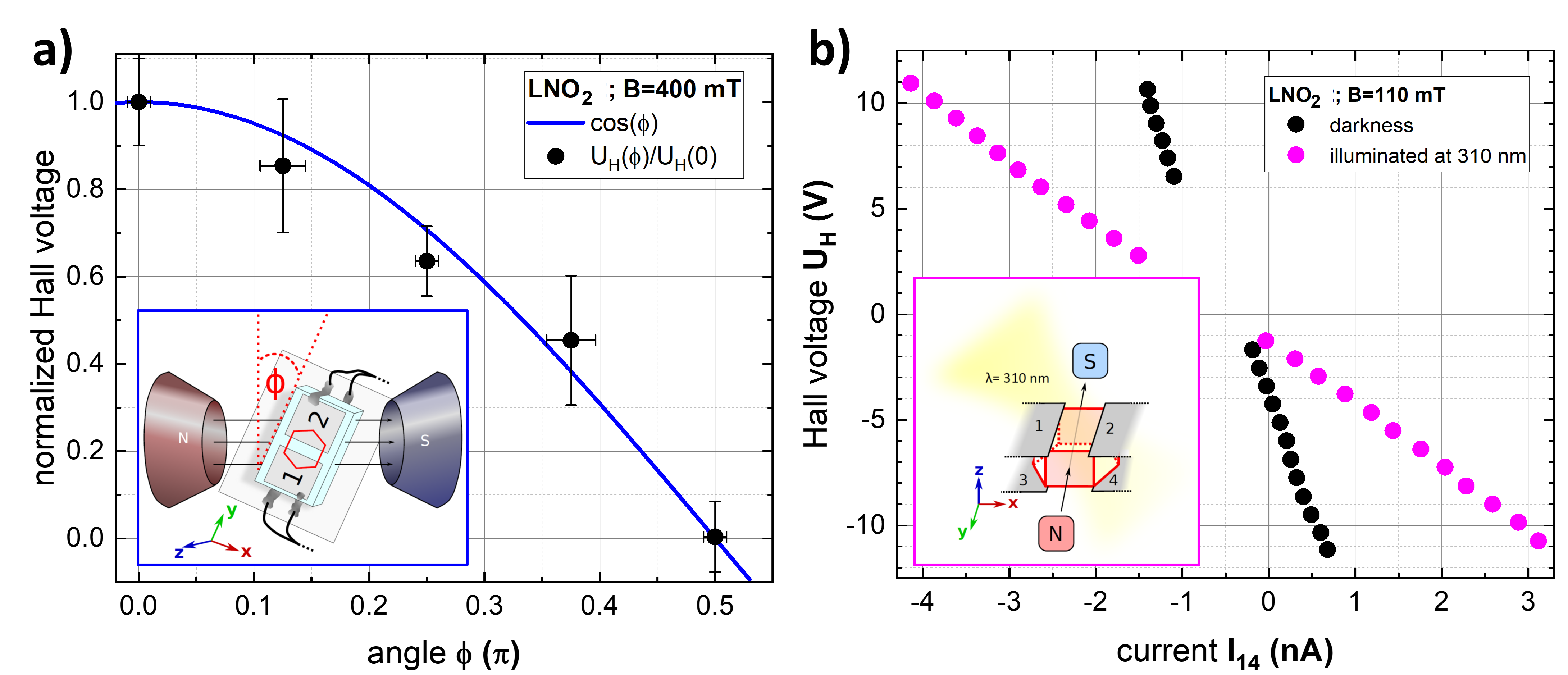}
\caption{\label{fig:3} (a) The DW of sample LNO$_2$ was exemplarily rotated in the $B$-field of the electromagnet, with the angle $\Phi$ being varied between 0 and 90$\symbol{23}$ and the corresponding Hall voltages measured, as sketched in the inset. As expected, the Hall voltage $U_H$ decreases when changing $\Phi$ from 0$\symbol{23}$ to 90$\symbol{23}$ and the ratio $U_H/U_{H,0}$ with $U_{H,0}$ being the Hall voltage at 0$\symbol{23}$ (i.e., with the magnetic field vector being aligned perpendicular to both the CDW and the injected current) follows the theoretically predicted cosine function very clearly. All data points were acquired with a constant magnetic field of 400~mT. (b) Impact of super-bandgap illumination on the Hall transport for sample LNO$_2$, which was placed in a constant field of $B$ = 110~mT supplied by a permanent magnet and illuminated at a wavelength of 310~nm under a constant photon flux of 10$^{13}$s$^{-1}$. The Hall voltage $U_H=U_{23}$ was recorded as a function of the current $I=I_{14}$ for the two cases: (i) under illumination (purple data points) and (ii) in the dark (black data points). Both data sets show a linear behavior, however, with a significantly increased response when illuminating the DW, then generating additional electron-hole pairs. }
\end{figure*}

Now, to verify that a true Hall voltage and not a parasitic quantity is measured, the angular dependence $U_H(\Phi)$ was recorded next, by mounting sample LNO$_2$ on a rotation table inside the electromagnet. Here, $\Phi$ denotes the angle between magnetic field vector $B$ and the plane of carrier transport, varying between 0$\symbol{23}$ and 90$\symbol{23}$ in a cosine fashion [see inset of Fig.~\ref{fig:3}(a)]. $U_H(\Phi)$ was then recorded for a fixed $B$-field of 400~mT at 5 different $\Phi$ values including 0$\symbol{23}$ and 90$\symbol{23}$. The data is plotted in Fig.~\ref{fig:3}(a), and clearly shows the expected Hall-voltage behavior, especially also recording $U_H(\Phi = 90 \symbol{23})$ = 0. 

\subsection{Hall Voltage under Super-Bandgap Illumination}

A second independent integrity experiment, carried out also only with specimen LNO$_2$, investigated the influence of super-bandgap illumination on the DW current, which is expected to significantly enlarge the sheet carrier density $n_{2D}$ by generating electron-hole pairs that then must decrease $R_h$ and $U_H$. For this complementary test the DW of LNO$_2$ was placed in a 110-mT field of a permanent magnet with $\Phi = 0 \symbol{23}$, and then illuminated at a 310~nm wavelength, which corresponds to the optical band gap of $E_g=4.0$~eV for bulk 5-mol\% Mg-doped LiNbO$_3$~\cite{sch12}. A 1000-W Xe arc lamp coupled into a grating monochromator (Cornerstone 260 by Oriel Instruments) and focused onto the whole 5$\times$10~mm$^2$ sample area served as the photo-exciting light source applying a constant photon flux of 10$^{13}$~s$^{-1}$ [see sketch in Fig.~\ref{fig:3}(b)]. Using this setup, $U_H$ was then acquired as a function of DW-current $I$ both with/without light. The data are displayed in Fig.~\ref{fig:3}(b). 

\subsection{Theoretical Background for the Extraction of Sheet Carrier Densities and Hall Mobilities}

Prior to discussing the results of these experiments, we briefly summarize here the mathematical background that is needed to extract both the charge carrier densities  $n_{2D}$ and the Hall mobilities $\mu_{H}$ from our experimental data. The full discussion was explained step-by-step in Beccard~\emph{et~al.}~\cite{bec22}, where we had already underlined the indispensable necessity for offset and error corrections in vdP experiments as outlined by Werner~\cite{wer17b}. To interpret our Hall data, we assume (A) that the DW current is established by electrons as the majority charge carriers. That the majority carrier are negative was derived from a preliminary experiment testing simply the polarity of the Hall voltage. The assumption that these carriers are very probably electrons is derived from two recent studies: first, via in-situ strain experiments for a set of equivalently prepared DWs based on crystal pieces of LNO wafers from the same manufacturer~\cite{sin22} and which can be understood due to their superior mobility compared to holes and the positive bound charges at a head-to-head (H2H)-DW requiring negative screening charges \cite{wer17,xia18}; and second, via temperature-dependent DWC measurements revealing activation energies in the range of 100-250~meV~\cite{zah23} and thus pointing towards electron-polaron mediated hopping with ionic transport being very improbable. And (B), we assume that the magnetic field stands perpendicular to the conductive layer, i.e., the CDW. Then, the Hall voltage $U_H$ follows~\cite{Schroder2006,lun09} the relationship $U_H=\frac{I\cdot B}{q\cdot n\cdot d}$ with $B$ the absolute value of the magnetic field, $I$ the current driven through the conducting DW layer (in particular: $I=I_{14}$), $q$ the elementary charge, $d$ the DW width, and $n$ the 3D charge carrier density. Introducing the 2D charge carrier density $n_{2D}$, also referred to as sheet carrier density, with $n_{2D}=n\cdot d$ and the Hall resistance $R_h=U_H/I=U_{23}/I_{14}$, the former can be extracted from the slope of the (measured) expected linear $R_h$-vs.-B dependence as:
 
\begin{equation}
\label{eq:2D_carrier_density}
    R_h \propto \frac{B}{q\cdot n_{2D}} \quad .
\end{equation}

\noindent For the subsequent calculation of the Hall mobility $\mu_H$ of the majority charge carriers, we employ the relation:

\begin{equation}
\label{eq:Hall_mobility}
    \mu_H=\frac{1}{q\cdot n_{2D}\cdot R_S} \quad .
\end{equation}

\noindent Note that the Hall mobility $\mu_H$ is linked to the "actual" mobility $\mu$ via the Hall factor $R_h$ as $\mu_H = R_h \cdot \mu$. The Hall factor, which depends on internal scattering mechanisms, is not known in most practical cases and commonly assumed to be unity~\cite{cam16}. Furthermore, $R_S$ is the sheet resistance of the CDW, which is readily obtained by solving van der Pauw's equation numerically~\cite{lun09}:

\begin{equation}
\label{eq:vdP_sheet_resistance}
    \exp{\left(-\frac{\pi}{R_S}R_{13,42}\right)}+\exp{\left(-\frac{\pi}{R_S}R_{34,21}\right)}=1 \quad .
\end{equation}

\noindent $R_{13,42}=U_{42}/I_{13}$ and $R_{34,21}=U_{21}/I_{34}$ are easily extracted from corresponding current injection and voltage measurements between the respective contacts. The values measured here are all listed in SI-table~S3.

\section{Results and Discussion}

As key findings of our study, the measured relationship between $R_h=U_H/I$ and the absolute value of the magnetic field $B$ normal to the DW is depicted for both DWs in Fig.~\ref{fig:2}. The corresponding raw data before averaging over both field directions can be found in the SI-Fig.~S5, while an additional averaging over the two possible current directions as in refs.~\cite{wer17b,bec22} could not be realized due to the rectifying character of the $U_{23}$-vs-$I_{14}$ curves. According to eq.~(\ref{eq:2D_carrier_density}), rewritten as $R_h \propto b\cdot B$ with the slope $b=\frac{1}{q \cdot n_{2D}}$, very clear and linear dependencies of the Hall resistance $R_h$ vs. magnetic field $B$ curves are indeed observed experimentally here for the DWs in both LNO samples. The corresponding slopes were extracted from the linear fits. They read as $b_1 = (31 \pm 3)\cdot 10^9~\Omega T^{-1}$ as well as  $b_2 = (21 \pm 3)\cdot 10^8~\Omega T^{-1}$ for samples LNO$_1$ and LNO$_2$, respectively. Table~\ref{tab:summary_calculations} summarizes the sheet carrier densities $n_{2D}$ as extracted from these slopes. The numerical values for $n_{2D}$ cover two orders of magnitude for the two specimen, with LNO$_2$ showing the larger $n_{2D}$ value of 3$\times$10$^5$cm$^{-2}$, which is two orders of magnitude more than recently observed in conductive BaTiO$_3$ DWs, and moreover, a very reasonable value for a 2D electronic system~\cite{bec22}. Reconsidering the findings of the CSHG imaging [cf.~Fig.~\ref{fig:1}(a),(c)] the $n_{2D}$-discrepancy between LNO$_1$ and LNO$_2$ is not a surprise, since the DW of LNO$_2$ shows a broader range of DW inclination angles. Consequently, a higher charge carrier reservoir in contrast to LNO$_1$ is obviously expected. In other words, the link between DW geometry, or more generally, the DW's real structure, and electrical performance, is reflected in the present Hall-effect results.

In principle, the 3D charge carrier density $n=n_{2D}/d$ can be calculated as well, provided the DW width $d$ is known. Nevertheless, for the two specimen here $d$ is unknown, but to tentatively estimate typical values we use the literature value of 174~pm, which was derived from transmission electron microscopy (TEM) by Gonnissen \textit{et al.}~\cite{gon16} on macroscopically non-inclined LNO domain walls, and receive carrier densities $n$ reading $1.15\cdot10^{12}~$cm$^{3}$ for LNO$_1$ and $17\cdot10^{12}~$cm$^{3}$ for LNO$_2$, respectively. However, as of now it is not clear, (i) whether the width of our conductivity-enhanced DWs here is comparable with the width of the DWs as measured by Gonnissen \emph{et al.}, and (ii) to what extent the transport channel is different in width compared to the width defined by the polarization change as measured via TEM, i.e., whether screening charges are trapped in a larger area. Therefore, the above values for $n$ should be seen as first "qualified guesses".\\

\begin{table}[h]
\caption{\label{tab:summary_calculations}%
Overview of the DW parameters, i.e., the sheet resistance $R_S$ (for their graphical determination see SI-tab.~S3 and SI-fig.~S4), the 2--dimensional charge carrier density $n_{2D}$, and the Hall mobility $\mu_H$, as extracted from  DW-confined Hall effect measurements using eqs.~(\ref{eq:2D_carrier_density})--(\ref{eq:vdP_sheet_resistance}).}
\begin{ruledtabular}
\begin{tabular}{lccc}
sample & $R_S$    & $n_{2D}$   &  $\mu_H$\\
        & ($10^{12}\Omega/\square$)  & (10$^3$/cm$^{2}$)  & (cm$^2$/Vs)\\
    \hline
LNO$_1$ & 5.6  & 20 $\pm$ 2     & 54 $\pm$ 5     \\
LNO$_2$ & 0.6  & 297 $\pm$ 36   & 35 $\pm$ 4   \\
\end{tabular}
\end{ruledtabular}
\end{table}

The second quantity evaluated from our four-point probe setup employing eqs.~(\ref{eq:Hall_mobility}) and (\ref{eq:vdP_sheet_resistance}), is the Hall mobility $\mu_H$. The extracted values are listed in Tab.~\ref{tab:summary_calculations} as well. In comparison to mobilities of LNO \emph{bulk} that have been reported~\cite{ohm76} to be 0.8~cm$^2$/Vs, the \emph{domain walls'} Hall mobilities found here are significantly (almost 2 orders of magnitude) higher, which is an expected and desirable result. On the other hand, the Hall mobilities in \emph{thin-film based} LNO domain walls, which were reported by Qian \emph{et~al.}~\cite{qia22} (337.30~cm$^2$/Vs) and McCluskey \emph{et~al.}~\cite{mccluskey_ultrahigh_2022} (around 3700~cm$^2$/Vs) are 1--2 orders of magnitude larger once more and it will be a future experimental challenge to investigate whether these ranges can be achieved within LNO single crystal DWs as well. One may question to what extent the preconditions that justify eq.~(\ref{eq:vdP_sheet_resistance}), are satisfied for the DWs of the current study. In an ideal case, the conducting sheet in the van-der-Pauw configuration has to be homogeneous, isotropic, uniform in thickness, without holes, and the contacts have to be point contacts  at the perimeter. Reconsidering the CSHG-microscopy images showing a significant range of inclination angles and a tendency towards spike domain formation at least for sample LNO$_2$ and the fact that the LNO DWs are more tube-like rather than forming a 2D sheet,
it appears to be necessary to estimate these errors as induced by these non-idealities:

\noindent In the simplest case and as indicated in Fig.~\ref{fig:1}(b,d), the Hall transport in samples LNO$_1$ and LNO$_2$ can be assumed as a parallel conduction through two identical DW sheets, which would mean a factor of 2 between the measured current and the current flowing through one of the two sheets, i.e., in turn, $n_{2D}$ would be diminished by a factor of 2. Accounting for the more complex real structure of the CDWs in samples LNO$_1$ and LNO$_2$ [see Fig.~\ref{fig:1}(a,c)] we accomplished resistor-network simulations as described in detail earlier~\cite{wol18} and briefly explicated in SI-section~D, with the rather similar result that the determined charge carrier densities are on the same order of magnitude needing correction factors of 0.51 and 0.65 for sample LNO$_1$ and LNO$_2$, respectively (see SI-sec.~D). Note that the other extreme case, i.e., a strongly asymmetric conductivity of the two parallel DW sheets, would not need such a correction, since the measurement would reflect mainly the "Hall-behavior" of the highly conductive DW. Against the background that in a Hall scenario primarily the \emph{order of magnitude} of carrier densities and mobilities are of interest, correction factors between 0.5 and 1 are fully acceptable in any case. 

Nevertheless, the non-ideal fitting to the van-der-Pauw restrictions gave further motivation to test the integrity of the approach by additional Hall-effect measurements:

In a first supporting experiment we recorded the dependence of $U_H$ on the angle $\Phi$ enclosed between magnetic field $B$ and conducting sheet as exemplarily illustrated in Fig.~\ref{fig:3}(a) for sample LNO$_2$, i.e., the DW with the largest charge carrier density and the lowest sheet resistance. As seen from the plot of the normalized Hall voltage $U_H/U_{H,0}$ versus angle $\Phi$, the expected cosine behavior is convincingly reproduced by the measured data.

In a second additional experiment illustrated in Fig.~\ref{fig:3}(b) performed again with the DW of specimen LNO$_2$, we studied whether super-bandgap illumination at 310~nm, which corresponds to the optical bandgap of Mg-doped LiNbO$_3$ of 4.0~eV, might impact the Hall voltage. In fact, the earliest experiments focusing on measuring the DW conductivity in LNO had already qualitatively reported that super-bandgap light strongly enhances the charge carrier density at DWs~\cite{sch12}. Indeed, the slope of the $U_H$-vs.-$I$ decreases under UV illumination, as shown in Fig.~\ref{fig:3}(b), indicating a decrease in $R_h$ and hence an increase in the sheet carrier density $n_{2D}$, as follows from eq.~(\ref{eq:2D_carrier_density}). The numerical evaluation of the slope via linear curve fitting yields a significant increase of the carrier density by a factor of 4--5 under super-bandgap excitation with a rather moderate photon flux. In the control experiment where a monodomain z-cut LiNbO$_3$ bulk crystal (i.e., without hexagonal domain structures) was covered with the same electrode configuration and tested in the same type of measurement scenario, a bulk photo-Hall effect could be excluded, since no functional relationsship between the driving current $I_{14}$ and the Hall voltage $U_{23}$ apart from noise was observed [SI-Fig.~S6(b)]. Thus, we state that the proposed Hall-effect measurement setup is also efficient for investigating the photo-induced DW-confined transport behavior, which clearly opens up the box to implement DW-based devices for nano-optoelectronic applications as well.  
\section{Conclusion}

In summary, two exemplary conductive ferroelectric domain walls, engineered into a z-cut 200-\textmu m-thick 5-mol\% MgO-doped LiNbO$_3$ single crystal, completely penetrating the latter, and shaped like hexagonal tubes with, however, different microscopic real structure, were macroscopically electrically connected with four Cr electrodes (two on the z$^+$ and two on the z$^-$ surface) in order to set up  a van-der-Pauw four-point-probe geometry for Hall probing. This setting in turn was employed for measuring the Hall resistance $R_h$ as a function of the magnetic field $B$ applied perpendicular to two of the six side sheets of the hexagonally-shaped DW tubes, as well as the sheet resistance $R_s$ of the DWs, which finally allowed us to extract two characteristic key quantities for such low-dimensional electronic systems, i.e., the 2D charge carrier density $n_{2D}$ found to be in the range 20...300$\cdot$10$^2$cm$^{-2}$ and the Hall mobility $\mu_H$ (extracted as 54 and 35~cm$^2$(Vs)$^{-1}$ for the two samples, respectively, with a reasonable error around 10\%). The validity of these numbers was further tested through angle- and illumination dependent Hall-voltage recordings, which both showed the expected behavior. Moreover, we employed resistor-network simulations to calculate correction factors for $n_{2D}$ and $\mu_h$, due to the parallel junction formed by the two CDWs, which were in the range of 0.5 and thus did not change the order of magnitude of the two quantities. Thus we propose that macroscopic Hall-effect analysis, as applied here, provides a robust and versatile method for the comparative quantification of the electrical performance of conductive domain walls in both LNO and many other materials. Moreover, photo-induced Hall measurements might gain even more interest, especially also for realizing nano-optoelectronic circuits.

\section*{Acknowledgements}

We acknowledge financial support by the Deutsche Forschungsgemeinschaft (DFG) through joint DFG--ANR project TOPELEC (EN~434/41-1 and ANR-18-CE92-0052-1), the CRC~1415 (ID: 417590517) the FOR~5044 (ID:
426703838; \url{https://www.for5044.de}) as well as through the W\"urzburg-Dresden Cluster of Excellence on "Complexity and Topology in Quantum Matter" - ct.qmat (EXC 2147, ID: 39085490). This work was supported by the Light Microscopy Facility, a Core Facility of  the CMCB Technology Platform at TU Dresden.

\section*{Data Availability}

The data that support the findings of this study are available
from the corresponding author upon reasonable request.

\bibliography{aipsamp}


\setcounter{table}{0}
\renewcommand{\thetable}{S\arabic{table}}  

\setcounter{page}{1}
\renewcommand{\thepage}{S\arabic{page}} 

\setcounter{equation}{0}
\renewcommand{\theequation}{S.\arabic{equation}} 

\setcounter{figure}{0}
\renewcommand{\thefigure}{S\arabic{figure}}

\setcounter{section}{0}

\onecolumngrid


\vspace{1cm}
\hrule

\section*{Supplementary Information}

\subsection{Compilation of current-voltage (IV) characteristics and derived DW resistances}

Here, we assemble a number of current-voltage curves, all of them collected with a Keithley~6517B electrometer, of the two different LNO DWs under investigation in order to complete the overall picture on them.

\begin{figure*}[!h]
\includegraphics[width=0.8\textwidth]{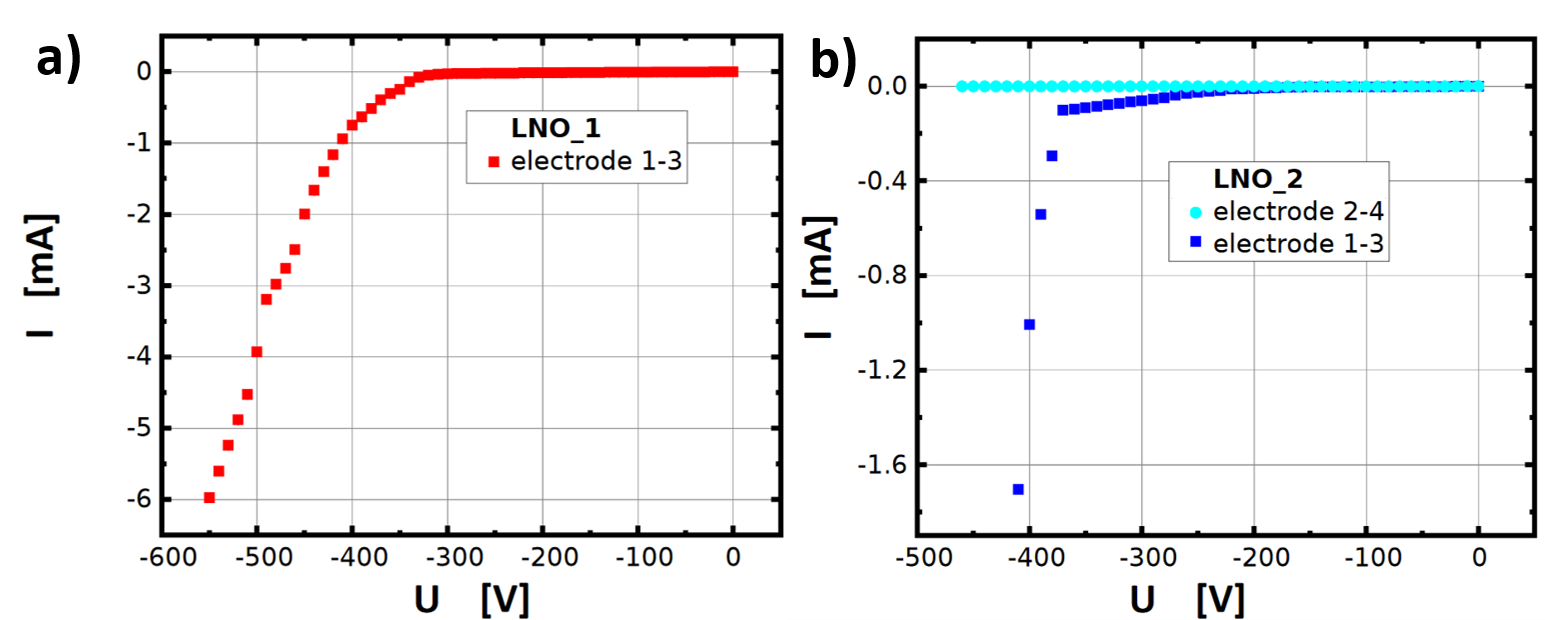}
\caption{\label{fig:IV_enhancement} Current-voltage characteristics logged \emph{during} the high-voltage biasing of the hexagonal LNO domain walls in order to increase the domain wall inclination and thus the DW conductivity according to the protocol of Godau et al.~\cite{god17}. a) In the case of LNO$_1$ only one side, i.e., one top-bottom electrode pair (1/3) was used. b) In sample LNO$_2$ both, the top-bottom pairs 1/3 and 2/4 were treated, whereas only in the first case a significant current flow could be induced. However, as SHG images (cf.~Fig.~1(c) of the main text) show, both sides of the hexagonal domain wall showed higher average inclinations with respect to the polar axis than in the case of LNO$_1$. The resulting after-conductivity-enhancement IV-curves measured in both z-direction and diagonally are shown in the following figures~\ref{fig:IV_z_dir} and \ref{fig:IV_diag}, respectively.}
\end{figure*}

First of all, the current-voltage characteristics during the high-voltage ramping treatment according to our protocol published earlier~\cite{god17} (using the voltage source of the Keithley~6517B) is shown in SI-Fig.~\ref{fig:IV_enhancement}. The corresponding final electrical fields $E_{13}$ and $E_{24}$ (the latter was applied only in case of LNO$_2$) are listed in SI-tab.~\ref{tab:LNO_parameters}. The voltage ramp velocity was 10~V per 2~s. 

Further, we acquired a number of standard low-voltage current-voltage characteristics between different combinations of electrodes in the range of $\pm$10~V. Therefore the measuring voltage was swept from $-$10~V to 10~V using the voltage source of a Keithley 6517B electrometer with waiting intervals of 2~s while subsequently measuring the current by the same device.

\begin{figure*}[!h]
\includegraphics[width=0.85\textwidth]{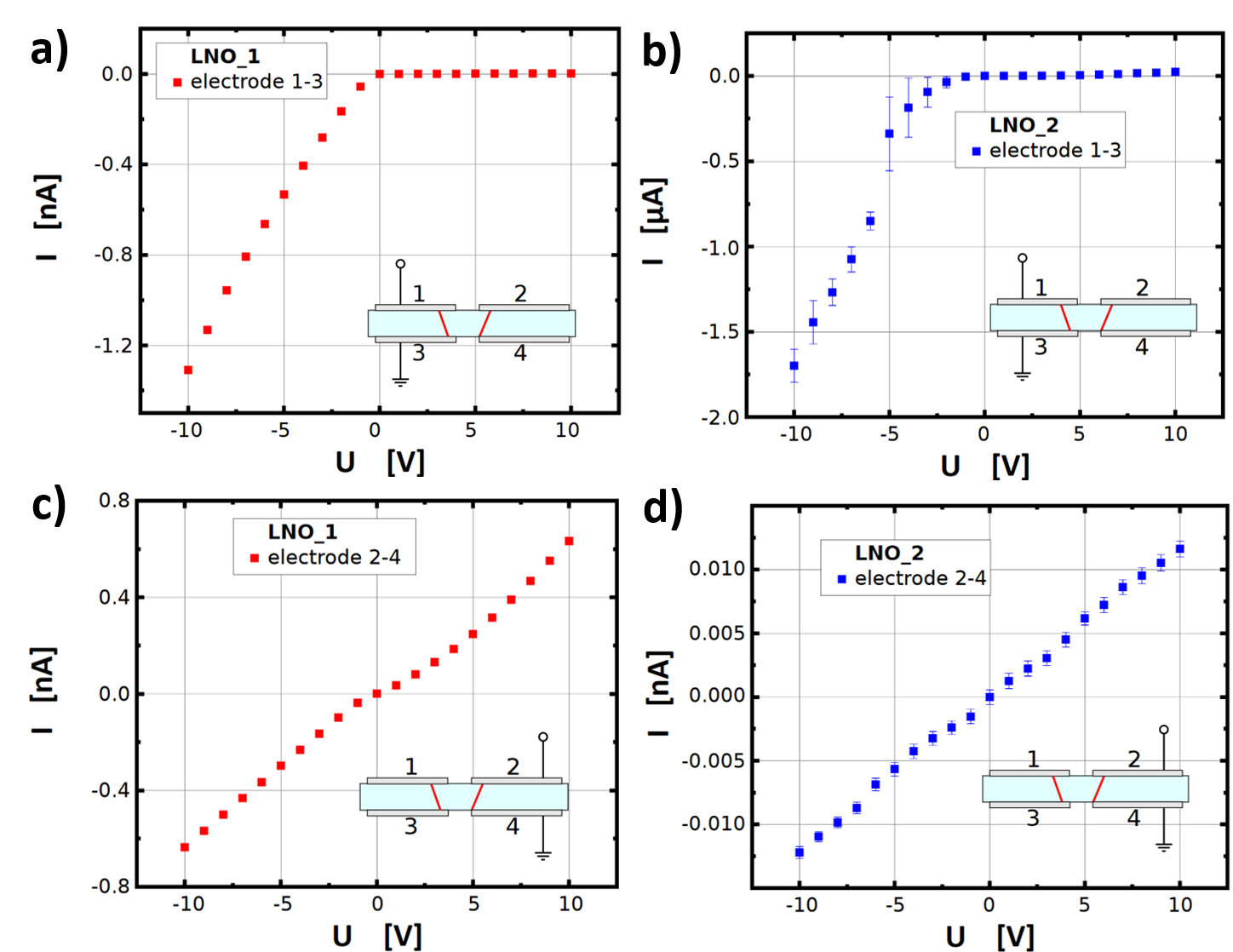}
\caption{\label{fig:IV_z_dir}Current-voltage characteristics in the $\pm$10~V range as measured along the z-direction between electrodes 1/3 and 2/4, respectively. Panels (a,c) and (b,d) depict the case of LNO$_1$ and LNO$_2$, respectively. There is a noticeable similarity between the two samples: The IV-characteristics between electrodes 1/3, which are the electrode pairs having shown a significant "enhancement current" (Fig.~\ref{fig:IV_enhancement}), are diode like, while the IV-curves for electrodes 2/4 are ohmic-like.}
\end{figure*}

SI-Fig.~\ref{fig:IV_z_dir} depicts the low-voltage IV-curves in z-direction (polarization direction) between the two electrode pairs. While the enhancement between electrodes 1 and 3 was obviously successful (with "conductivity-enhancement factors" $F_{13}$=6$\cdot 10^{3}$ and 5$\cdot 10^{6}$ for LNO$_1$ and LNO$_2$, respectively, cf.~SI-tab.~\ref{tab:LNO_parameters}) and led to diode-like characteristics [SI-Fig.~\ref{fig:IV_z_dir}(a) and (b)], the other side, i.e., the path between electrodes 2 and 4, exhibits ohmic behavior and a detectable but decisively lower conductivity enhancement ($F_{24}$=3$\cdot 10^{3}$ and 9$\cdot 10^{1}$) for LNO$_1$ and LNO$_2$, respectively. This is not surprising, since for LNO$_1$ no high-voltage ramp had been driven between this electrode pair intentionally, and in the case of LNO$_2$ the enhancement protocol was obviously not successful (cf. SI-Fig.~\ref{fig:IV_enhancement}(b) again). 

\begin{figure*}[!h]
\includegraphics[width=0.85\textwidth]{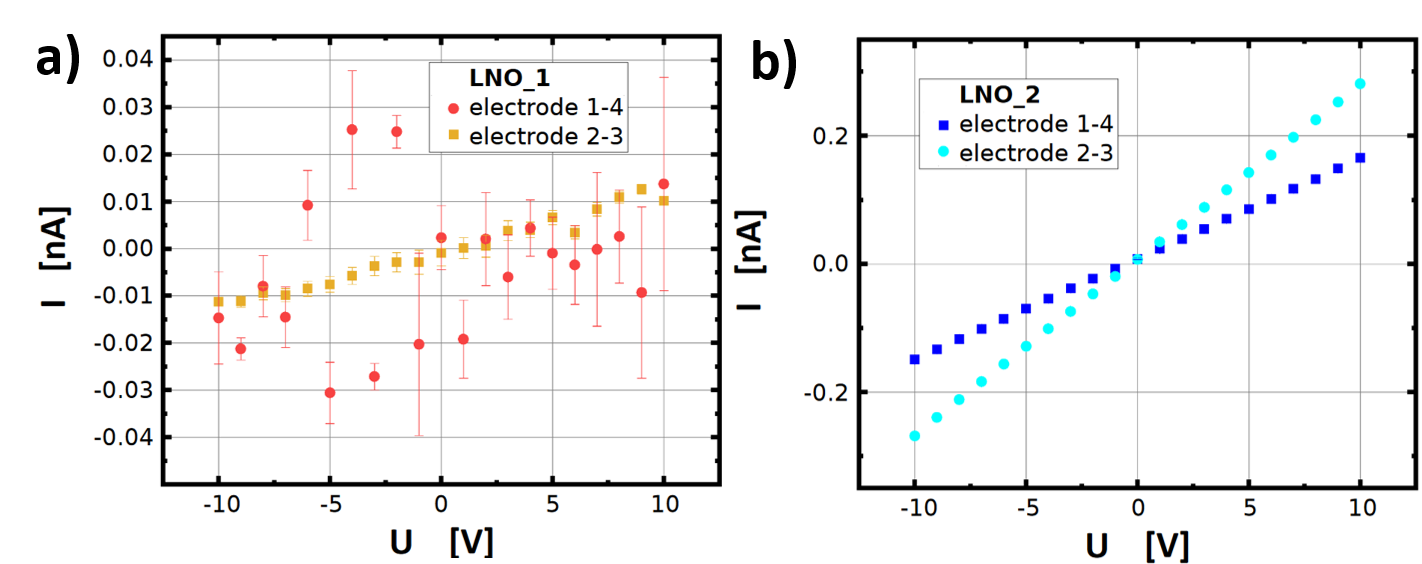}
\caption{\label{fig:IV_diag}Post-enhancement diagonal IV-characteristics for the DWs of samples (a) LNO$_1$ and (b) LNO$_2$. The corresponding DW resistances, which were extracted from the slopes after linear curve fitting, are listed in tab.~\ref{tab:Diagonal_resistances}.}
\end{figure*}

\begin{table}[!h]
\caption{\label{tab:LNO_parameters}
Overview of selected electrical parameters of the two LNO domain walls of the study: as-grown resistance in z-direction between electrodes 1/3, $R^{(0)}_{13}$; electrical field during the "enhancement" treatment, $E_{13}$; post-enhancement resistance in z-direction between electrodes 1/3: in forward direction -- $R_{13}^{(forw)}$, in reverse direction -- $R_{13}^{(rev)}$; conductivity-enhancement factor $F=R^{(0)}_{13}/R_{13}^{(forw)}$. For the electrode pair 2/4 we use an analogous nomenclature with $R_{24}$ covering the uniform since ohmic "post-enhancement" resistance.}
\renewcommand{\arraystretch}{1.5}
\begin{ruledtabular}
\begin{tabular}{lccccccccc}
   sample & $R_{13}^{(0)}$ ($\Omega$) & $E_{13}$ ($\frac{kV}{mm}$)& $R_{13}^{(forw)}$ ($\Omega$) & $F_{13}$ & $R_{13}^{(rev)}$ ($\Omega$) & $R_{24}^{(0)}$ ($\Omega$) & $E_{24}$ ($\frac{kV}{mm}$)& $R_{24}$ ($\Omega$) & $F_{24}$\\
\hline
LNO$_1$ &  71$\cdot 10^{12}$ & 2.75 & 11$\cdot 10^{9}$ &  6$\cdot 10^{3}$ & 0.3$\cdot 10^{9}$ & 57$\cdot 10^{12}$ & -- & 18$\cdot 10^{9}$ &  3$\cdot 10^{3}$\\
LNO$_2$ & 12$\cdot 10^{12}$& 2.05& 2.6$\cdot 10^{6}$ &  5$\cdot 10^{6}$ & 0.4$\cdot 10^{9}$ & 74$\cdot 10^{12}$& 2.30 & 833$\cdot 10^{9}$ &  9$\cdot 10^{1}$\\
\end{tabular}
\end{ruledtabular}
\end{table}
\renewcommand{\arraystretch}{1}

\begin{table}[!h]
\caption{\label{tab:Diagonal_resistances}%
Diagonal resistances $R_{14}$ and $R_{23}$ as measured in the "post-enhancement" state.}
\begin{minipage}{0.3\textwidth}
\renewcommand{\arraystretch}{1.5}
\begin{ruledtabular}
\begin{tabular}{lcc}
   sample & $R_{14}$ ($10^9\Omega$) & $R_{23}$ ($10^9\Omega$)\\
\hline
LNO$_1$ & 625 & 847 \\
LNO$_2$ & 50 & 33 \\
\end{tabular}
\end{ruledtabular}
\end{minipage}
\end{table}
\renewcommand{\arraystretch}{1}

\begin{table}[!h]
\caption{\label{tab:Raw_Data_sheet_resistance}%
Measured current and voltage values in van-der-Pauw configuration, as well as following resistance values $R_{13,42}$ and $R_{34,21}$ for the extraction of the sheet resistance $R_S$ (cf.~tab.~I of the main paper) according to eq.~(3) of the main text.}
\begin{minipage}{0.8\textwidth}
\renewcommand{\arraystretch}{1.5}
\begin{ruledtabular}
\begin{tabular}{lcccccc}
   sample & $R_{13,42}$ (10$^{12}\Omega$) & $U_{42}$ (V) & $I_{13}$ (10$^{-12}$A) & $R_{34,21}$ (10$^{12}\Omega$) & $U_{21}$ (V) & $I_{34}$ (10$^{-12}$A)\\
\hline
LNO$_1$ & 2 & 3.7 & 1.9 & 0.7 & 6.2 & 8.9 \\
LNO$_2$ & 0.18 & 0.5 & 2.8 & 0.1 & 0.3 & 2.8 \\
\end{tabular}
\end{ruledtabular}
\end{minipage}
\end{table}
\renewcommand{\arraystretch}{1}

\begin{figure*}[!h]
\includegraphics[width=0.75\textwidth]{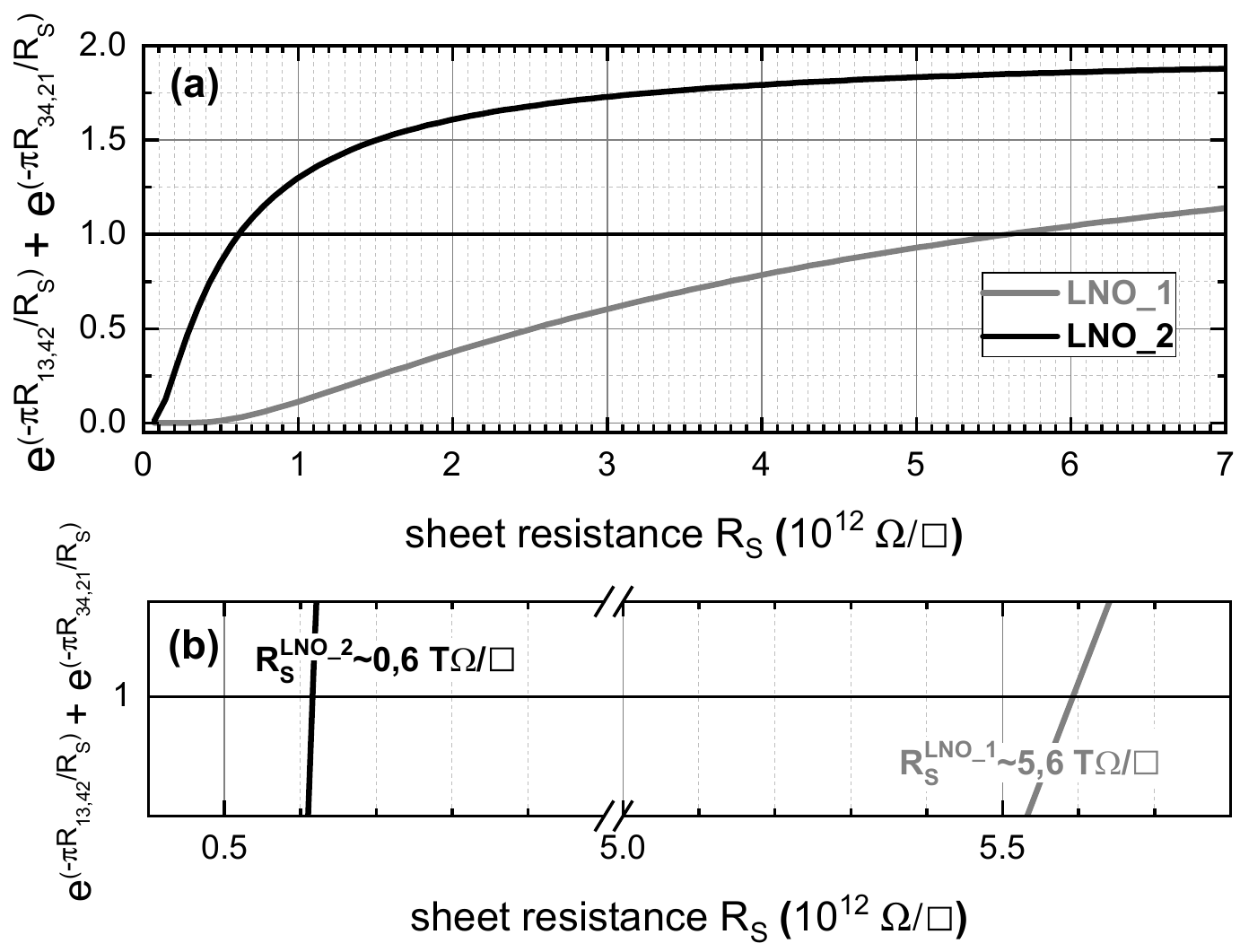}
\caption{\label{fig:vdP_sheet_resistances}Determination of the domain walls' sheet resistances $R_S$ from a graphical solution of eq.~(3) employing the measured resistance values $R_{13,42}$ and $R_{34,21}$ (cf.~SI-tab.~\ref{tab:Raw_Data_sheet_resistance}). Panel (b) is a close-up view of panel (a) in order to more clearly show the intersection points with the Y$=$1-line, hence numerically revealing the $R_S$ values. }
\end{figure*}

\clearpage

\subsection{Selected additional Hall-effect data}

\begin{figure*}[!h]
\includegraphics[width=0.75\textwidth]{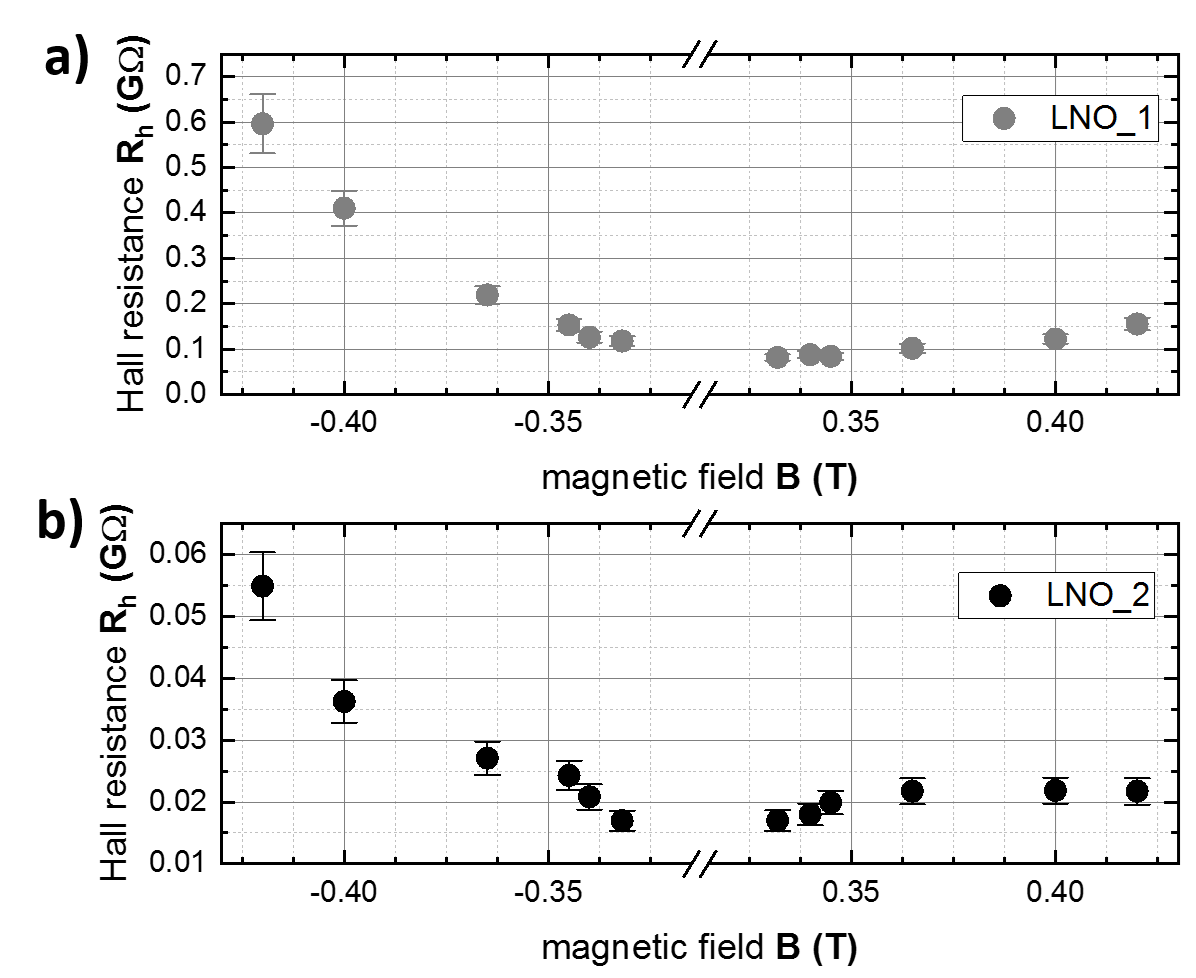}
\caption{\label{fig:RH_no_avg}$B$-field-direction-dependent Hall resistance for the cases of sample (a) LNO$_1$ and (b) LNO$_2$, which represent the "raw" data underlying Fig.~2 in the main text.}
\end{figure*}

\begin{figure*}[!h]
\includegraphics[width=0.7\textwidth]{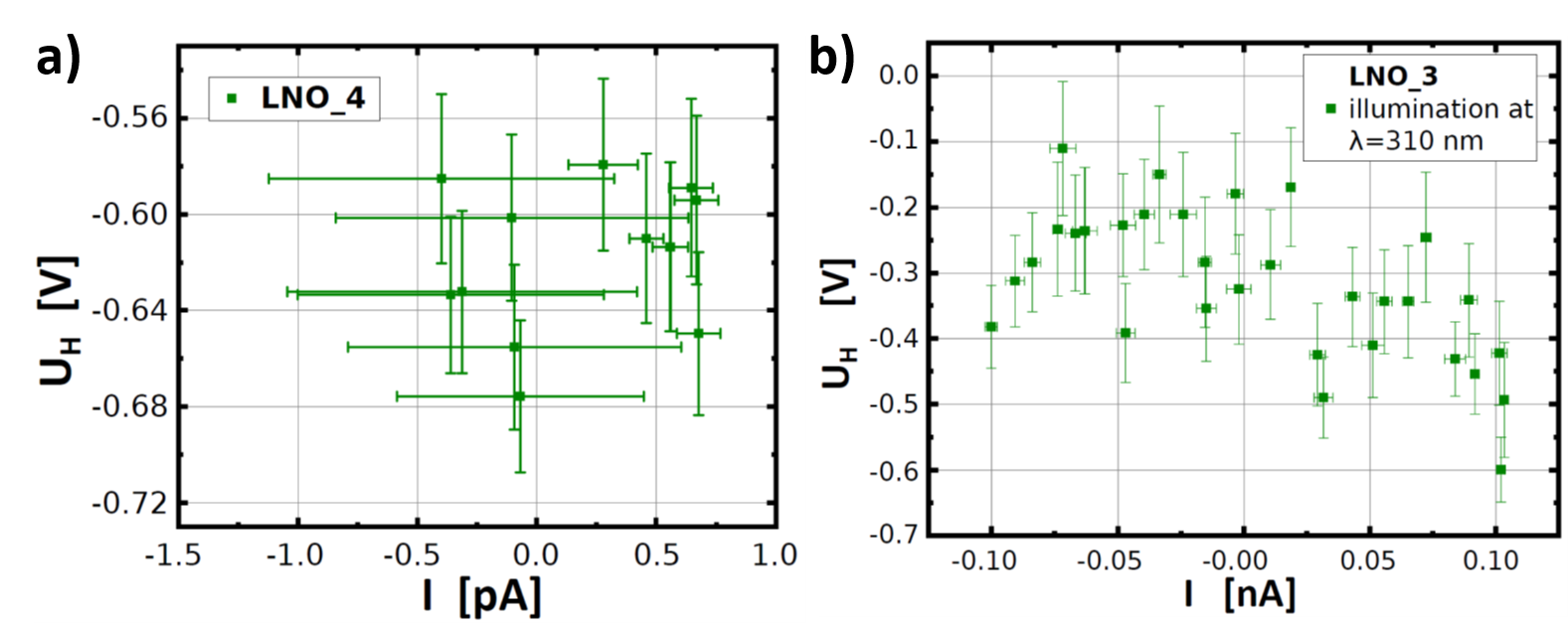}
\caption{\label{fig:reference_data}Reference data showing the Hall-voltage versus current characteristics of the two \textbf{LNO bulk monodomain crystals} LNO$_3$ and LNO$_4$, that contain no domain walls at all. The same 4-electrode configuration as for samples LNO$_1$ and LNO$_2$ was applied. Then all Hall measurements were carried out in exactly the same way, i.e. (a) in the dark under an exemplary $B$-field of 250~mT, and (b) under the 310-nm illumination in a $B$-field of 110~mT. Figs. (a,b) indicate no functional relation to any of the parameters (including also no bulk photo-Hall effect), and thus determine the noise level for the two experiments.}
\end{figure*}

\clearpage

\subsection{Supporting microscopic domain-wall images}

\begin{figure*}[!h]
\includegraphics[width=0.75\textwidth]{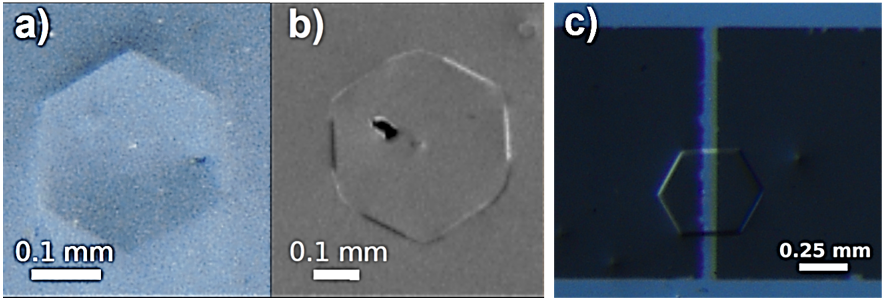}
\caption{\label{fig:polarization_microscopy}Polarization microscopy images of samples (a) LNO$_1$ and (b) LNO$_2$. (c) Polarization microscopy image (top view) of a hexagonal LNO domain also showing the two top vdP chromium electrodes (black rectangles). \\}
\end{figure*}

Complementary to the standard IV characterization, the domains were visualized by default via polarization microscopy [2D top view for both DWs, SI-figs.~\ref{fig:polarization_microscopy}(a) and (b)]. In order to illustrate the electrode geometry, a polarization microscopy image with both, a hexagonal LNO domain and the two top electrodes, is depicted in SI-Fig.~\ref{fig:polarization_microscopy}(c). 

\subsection{Random-Resistor-Network simulations}

Normally, in 4-point Hall measurements only a single conductive sheet is measured. However, due to the specific geometry of the hexagonal domains the samples of LNO$_1$ and LNO$_2$ contain two conductive sheets in a parallel configuration, which both contribute to the Hall voltage. As discussed in the main text, this can lead to overestimations of the actual carrier densities (and conversely underestimations of the mobilities), because current flows in both sheets as shown in Fig. S8. The actual influence depends on the relation of conductivity of the two parallel sheets. In this regard, two extreme cases can be distinguished:  (1) Both sheets are of equal conductivity. Hence, the applied current splits equally between both sheets leading to a overestimation of the charge carrier density and an underestimation of mobility. Hence, the extracted carrier densities need to be multiplied by a value of 0.5. Subsequently, if the mobilities are calculated in the uncorrected case, they will be overestimated by a factor of 2, respectively. (2) In contrast, one sheet might be much more conductive compared to the other one. Then, the more conductive sheet will dominate the transport measurement. Therefore, only this current will contribute to the measured Hall-voltage. In that case, no correction is necessary, i.e. the correction factor is 1 for both, as this represents the standard 4-point Hall configuration. \\
Any real structure of two parallel sheets will require a correction between these two extremes. In our experiment, we could not measure the conductivity independently. However, we can reconstruct the (local) conductivity of each sheet based on the 3D-SHG data and a previously developed and verified random-resistor-network (RRN) model \cite{wol18}. It is well established that the DW conductivity is proportional to the local DW inclination with respect to the polar axis. This can be measured with a resolution down to the µm in three dimensions with SHG microscopy \cite{kir19}. The resulting scans are shown in Fig. 1 in the main text, where the local inclination is color coded. For the two contributing sheets, the local inclination is converted into a map of ohmic resistors, which are each connected to four neighbors and the value of each resistor is propertional to the sine of the local inclination. The workflow is shown for one sheet in Fig.~S8. \\
The resulting resistor network of each sheet allows to calculate the local potential by numerically solving the Poisson equation. From the resulting potential distribution the current flowing in each node of the network is determined. Thus allowing the calculation of the ratio of current flowing through each sheet. More details on the algorithm can be found in the original work by Wolba at al.~\cite{wol18}. Since the charge carrier density is directly proportional to the amount of current flowing through the Hall sample, these ratios serve as correction factors for the measured charge carrier densities as well. Subsequently the comparison between measured current and the ”real”, i.e., simulated, current leads to a correction factor of 0.51 for LNO$_1$ and 0.65 for LNO$_2$, respectively, which are close to the ideal factor of 0.5 for two equally conductive sheets in a parallel configuration. The difference in the correction factors also mirrors the observed differences in local inclination angles and shows that LNO$_2$ is closer to a single sheet Hall-sample due to its inhomogeneously distributed inclination angles being more prevalent on one side than the other.

\begin{figure*}[ht]
\includegraphics[width=0.95\textwidth]{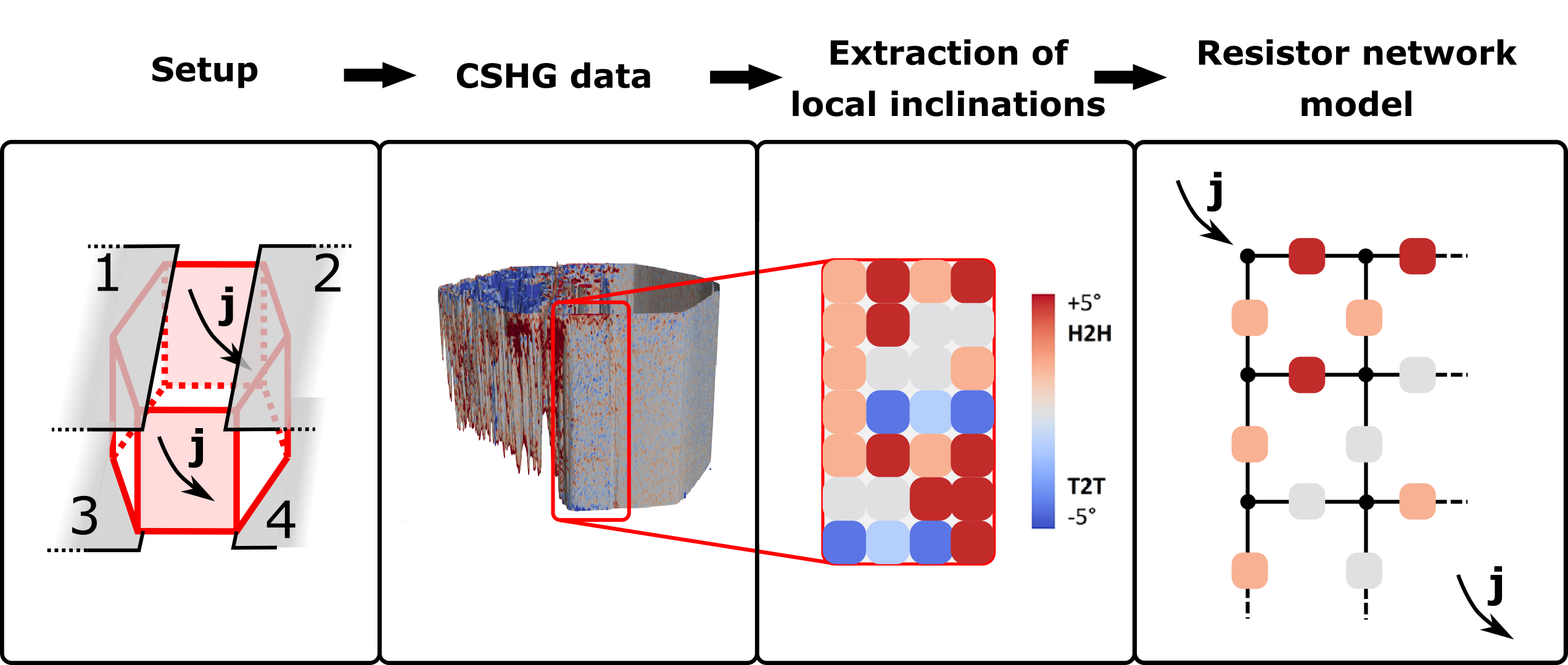}
\caption{\label{fig:resistor-network}Flowchart sketching the extraction of correction factors for both charge carrier density and Hall mobility, which becomes necessary due to the real structure of LNO DWs, which are not single sheets. Using second-harmonic-generation microscopy to determine the DWs local inclination angle. The inclination angles are mapped along two 2D planes and then used as the resistors in the resistor network.}
\end{figure*}

\end{document}